\documentclass[aps,prd,superscriptaddress,nofootinbib]{revtex4}
\usepackage[mathscr]{euscript}

\newcommand{\casimir}{\widetilde{\mathcal{I}}}
\usepackage{amsmath,amsfonts,amssymb,amsthm,amstext,amscd,graphicx}
\usepackage[all]{xy}
\usepackage[colorlinks=true]{hyperref}
\usepackage{amsmath}
\usepackage{bigints}

\def\be{\begin{equation}}
\def\ee{\end{equation}}
\def\arr{\begin{array}{rll}}
\def\ea{\end{array}}
\def\bea{\begin{eqnarray}}
\def\eea{\end{eqnarray}}

\def\N2{$N{=}2$}

\begin{document}

\title{Note on constants of motion in conformal mechanics associated with  near horizon extremal  Myers-Perry black holes}

\author{Hovhannes Demirchian}
\email{demhov@yahoo.com}
\affiliation{ V. Hambardzumyan Byurakan Astrophysical Observatory, Byurakan, 0213,  Armenia}
\begin{abstract} We investigate  dynamics of probe particles moving in the near-horizon limit of $(2N+1)$-dimensional extremal Myers-Perry black hole (in the cases of $N=3,4,5$) with arbitrary rotation parameters. Very recently it has been shown \cite{non-equal-general} that in the most general case with nonequal nonvanishing  rotational parameters the system  admits separation of variables in $N$-dimensional ellipsoidal coordinates. We wrote down the explicit expressions of Liouville integrals of motion, given in \cite{non-equal-general} in ellipsoidal coordinates,  in initial "Cartesian" coordinates in seven, nine and eleven dimensions, and found that these expressions hold in any dimension. Then, taking the limit where  all of the rotational parameters are equal, we reveal that each of  these $N-1$ integrals of motion  results in  the Hamiltonian of the spherical mechanics of a $(2N+1)$-dimensional MP black hole with equal nonvanishing  rotational parameters.
\end{abstract}
\maketitle


\setcounter{footnote}0
\section{Introduction}
Studies of near horizon geometries have great importance for observational and theoretical physics. Geodesics in this limit are associated with black hole accretions which might be the source of Very High Energy (VHE) gamma-ray bursts. Accretions around black holes can also be the key to the first direct observation of a black hole \cite{Falcke:1999pj} (e.g. with the Event Horizon Telescope). Near horizon geometries of extremal black holes are especially interesting. Extremal black holes have the lowest possible mass for the given charge or angular momentum. They also have vanishing surface gravity and vanishing Hawking temperature and hence their geometry is much simpler compared to other black holes of the same type.
The stationary extremal black holes in the near horizon  region have an isometry group $SO(2,1)$ \cite{bh,NHEG-general,NHEG-2}. Respectively, the particle dynamics on the near horizon extreme geometries possesses  dynamical conformal symmetry, i.e.  defines ``conformal mechanics" \cite{conformal-mechanics-BH-1,conformal-mechanics-BH-2, Anton-MP, Anton-1, GNS-1}.
Due to this symmetry one can perform canonical transformation under which the  Hamiltonian of the system  takes formally non-relativistic form \cite{conformal-mechanics-BH-2,GNS-1}
\be\label{nonrel}
H=\frac{p^2_R}{2}+\frac{2\mathcal{I}}{R^2},
\ee
where
$R=\sqrt{2K}$, $ p_R=\frac{2D}{\sqrt{2K}}$
are  the effective ``radius" and its canonical conjugate ``radial momentum", and $\mathcal{I}=HK-D^2$ is the Casimir of the conformal  algebra whose generators $H,D,K$ satisfy the relataions
\be
\{H,D\}=H, \quad \{H,K \}=2D, \quad \{D,K \} =K.
 \label{confalg}\ee
The Casimir  $\mathcal{I}$ depends on the $d-1$  ``angle-like" variables and their conjugate momenta which  commute with ``radial variables" $R,p_R$.  Hence, all  specific
properties of such systems are encoded in $\mathcal{I} $ which in turn may be viewed as the Hamiltonian of another associated system. Such associated systems have been investigated from  various viewpoints  in Refs. \cite{Armen-Tigran} where they were called ``angular (or spherical) mechanics".

In this paper we continue the study of the particle dynamics in  Near Horizon limit of Extremal Myers-Perry (NHEMP) black holes \cite{mp,bh,NHEG-general,NHEG-2}.

The case of EMP black holes with Equal nonvanishing Rotation Parameters (EMPERP) has been studied in \cite{Anton-MP,GNS-1},  where separability of variables of such a system in spherical coordinates  has been proven. Moreover, it has been shown that in the case of odd dimensions the spherical system is maximally superintegrable ($N$ dimensional system contains $2N-1$ functionally independent first integrals) and in even dimensions it lacks one integral of motion to be maximally superintegrable.

EMP black holes with Nonequal nonvanishing Rotation Parameters (EMPNRP) have been studied in \cite{non-equal-general} where it was shown that such systems admit separation of variables in  ellipsoidal coordinates, although they are not, seemingly,  superintegrable (in the case of equal parameters the symmetries of EMP black hole are enhanced). In this paper we continue studying EMPNRP and find explicit expressions for integrals of motion of its spherical mechanics in seven, nine and eleven dimensions, written in initial latitudial coordinates. Furthermore, we show that by simply requiring the rotation parameters to be equal in these expressions, one does not recover the first integrals of EMPERP. In such a special case all of the first integrals of the spherical mechanics of EMPNRP transform into the Hamiltonian of the spherical mechanics of EMPERP.

The paper is organized as follows.\\
In the Section \ref{sec:conformal_mechanics}, following \cite{non-equal-general}, we present the the details of conformal mechanics associated with EMPNRP, including the expression of
 the Hamiltonian of the spherical mechanics in spherical and elliptic coordinates.

 In the Section \ref{sec:integrals} the expression for first integrals in ellipsoidal and lattitudial coordinates will be derived for the systems related with  seven-, nine- and eleven- dimensional MPblack holes.
  
  In the last section \ref{sec:conclusion} we give the conclusion of the results.

\section{NHEMP Conformal mechanics}
\label{sec:conformal_mechanics}
Myers-Perry black holes \cite{mp} are $d$ dimensional, asymptotic flat, Einstein vacuum solutions. For $d=2N+1$ case these solutions come with $N$ angular momentum/velocity parameters $a_i$  and a mass parameter. In the extremal case the mass parameter is given in terms of the angular momentum parameters. The NHEMP metric in the appropriate parametrization takes the form
\be
\frac{ds^2}{r^2_H}=A(x)\left(-r^2d\tau^2+\frac{dr^2}{r^2}\right)+\sum_{i=1}^N dx_idx_i+
\sum_{i,j=1}^N\tilde{\gamma}_{ij}x_i x_j D\varphi^iD\varphi^j,\qquad D\varphi^i\equiv d\varphi^i+k^ird\tau
\label{m2}\ee
where
\begin{subequations}\begin{align}
A(x) =&\frac{\sum_{i=1}^N x^2_i/m^2_i}{1-8\sum_{i<j}(m_i m_j)^{-1}},\label{A}\\
\tilde{\gamma}_{ij}=&\delta_{ij}+ \frac{1}{\sum_l x_l^2/m^2_l}\frac{\sqrt{m_i-1}x_i}{m_i}  \frac{\sqrt{m_j-1}x_j}{m_j},
\label{gamma}\\
k^i&=2\frac{\sqrt{m_i-1}/m_i}{1-8\sum_{l<n}(m_lm_n)^{-1}},\label{k}
\end{align}\end{subequations}
 The horizon radius of the black hole  $r_H$  is     defined by the maximal value of the solution of equation
\be\label{ai}
\sum_i\frac{r^2_H}{r^2_H+a^2_i}=1,
\ee
where $a_i$ are rotational parameters.
The parameters $m_i$ are related with $a_i$ and $r_H$  as follows
\be\label{m-i-a-i}
m_i=\frac{{r^2_H+a^2_i}}{r^2_H}\geq 1.
\ee
Besides,  parameters $m_i\geq 1$ and coordinates  $0<x_i\leq \sqrt{m_i} $  obey the conditions
\be
\sum_{i=1}^N \frac{x^2_i}{m_i}=1,\quad \sum_{i=1}^N\frac 1m_{i}=1.\label{ell}
\ee

In the above notation the mass-shell equation for a particle of mass $m_0$ moving in the background metrics \eqref{m2}, $m^2_0=\sum_{A,B=1}^{2N+1}g^{AB}p_A p_B$,
leads to the following expression for the generators of conformal algebra
\begin{gather}
H=p_0= r\left(\sqrt{ \mathcal{I}(p_a,x_a, p_{\varphi_i})+ \left( \sum_i k_ip_{\varphi_i}\right)^2 +(rp_r)^2}+ \sum_i k_ip_{\varphi_i}\right),\\
 D=r p_r,\qquad
K=\frac{1}{r} \left(\sqrt{ \mathcal{I}(p_a,x_a, p_{\varphi_i})+ \left( \sum_i k_ip_{\varphi_i}\right)^2 +(rp_r)^2}- \sum_i k_ip_{\varphi_i}
 \right),
\label{ho}
\end{gather}
where the momenta $p_a, p_{\varphi_i}, p_r$ are conjugated to $x_a, \varphi_i, r$ with the canonical Poisson brackets
\be\label{Poisson-bracket}
\{p_a,x_b\}=\delta_{ab}, \qquad \{p_{\varphi_i},\varphi_j\}=\delta_{ij},\qquad\{p_r, r\}=1.
\ee

Here $\mathcal{I}$ is the Casimir element of conformal algebra
\begin{gather}
\mathcal{I}
=\frac{\sum_{i=1}^N x^2_i/m^2_i}{1-8\sum_{i<j}(m_i m_j)^{-1}}\left[\sum_a h^{ab}p_ap_b
+\sum_i\frac{p^2_{\varphi_i}}{x^2_i} +g_0
\right]- \mathcal{I}_0  ,
\label{L}
\end{gather}
where
\be
h^{ab}=\delta^{ab}-\frac{1}{\sum_i x^2_i/m^2_i}\frac{x_a}{m_a} \frac{x_b}{m_b},\qquad {\tilde\gamma}^{ij}=\delta^{ij} - \frac{x_i\sqrt{m_i-1}}{m_i} \frac{x_j\sqrt{m_j-1}}{m_j} ,
\ee
 and
\be\label{consts}
 g_0=\left(\sum_{i=1}^N \frac{\sqrt{m_i-1}p_{\varphi_i}}{m_i}\right)^2+m^2_0r^2_H,
 \qquad
 \mathcal{I}_0=4\left(\sum_i \frac{\sqrt{m_i-1}p_{\varphi_i}}{m^2_i}\right)^2.
\ee
Hence, the Hamiltonian \eqref{ho} can be represented in formally nonrelativistic form \eqref{nonrel},
with $R=\sqrt{2K}$ and $p_R=D/\sqrt{2K}$.

The Casimir   $\mathcal{I}$ \eqref{L} is at most quadratic in momenta canonically conjugate to the remaining angular variables and it can conveniently be viewed as the Hamiltonian of a reduced ``angular/spherical mechanics"  describing motion of particle on some curved background.
Since  the azimuthal angular variables $\varphi^i$ are cyclic, corresponding conjugate momenta $p_{\varphi_i}$ are constants of motion.
 We then remain with a reduced $N-1$ dimensional system described by Hamiltonian \eqref{L} and $x_a$ variables and their conjugate momenta.

 When the rotational parameters coincide, $m_i={N}$, the Hamiltonian of probe particle  reduces to the system on sphere and admits separation of variables in spherical coordinates \cite{GNS-1}. It can be interpreted as a singular spherical (Higgs) oscillator and is superintegrable system , i. e. possesses maximal number of functionally independent integrals of motion, $2(N-1)-1=2N-3$.

Other limiting case when neither of $m_i$'s  are equal analyzed in \cite{non-equal-general} admits separation of variables in the N-dimensional ellipsoidal coordinates $\lambda_i$.
\be
x^2_i=(m_i-\lambda_i)\prod_{j=1, j\neq i}^{N}\frac{m_i-\lambda_j}{m_i-m_j},
\label{xN}\ee
where $\lambda_N  < m_N  <  \ldots < \lambda_2  < m_2  < \lambda_1 < m_1$. To resolve the equation \eqref{ell} one should choose $\lambda_N=0$.

In these coordinates
the angular  Hamiltonian \eqref{L} reads
\be
\tilde{\mathcal{I}}=\lambda_1\ldots\lambda_{N-1}\left[ - \sum_a\frac{{4\prod_{i=1}^{N}(m_i-\lambda_a) }\pi^2_a}{\lambda_a\prod_{b=1,a\ne b}^{N-1}(\lambda_b-\lambda_a)}
+\sum_{i=1}^N\frac{{g}^2_i}{\prod_{a=1}^{N-1}(m_i-\lambda_a)} +g_0\right],
\label{24}\ee
 where
\be
	{ g}^2_i =\frac{p^2_{\varphi_i}}{m_i}\prod_{j=1,j\neq i}^{N} (m_i-m_j),\qquad \tilde{\mathcal{I}}\equiv  \left({\mathcal{I}+\mathcal{I}_0}\right){(1-8\sum_{i<j}(m_i m_j)^{-1})\prod_{i=1}^{N}m_i}
\ee
and $\{\pi_a,\lambda_b\}=\delta_{ab}$, $\{p_{\varphi_i},\varphi_j\}=\delta_{ij}$.

Our goal is to study the integrals of motion in this system.
\section{Integrals of motion}\label{sec:integrals}

The expressions for commuting integrals of motion $F_a$ can be found from\footnote{Note that $R_a\lambda_a\rightarrow R_a$ and $\nu_a \rightarrow F_{a+1}$ replacements have been assumed in the  current paper compared to \cite{non-equal-general}.} \cite{non-equal-general}
\be
\sum_{\alpha =1}^{N-1}F_{\alpha}\lambda^{\alpha-1}_a=R_a(\pi_a,\lambda_a)\qquad
\Longleftrightarrow \qquad
 \begin{pmatrix}
1 & \lambda_1 & \lambda_1^2 & \cdots & \lambda_1^{N-2}
\\
1 & \lambda_2 & \lambda_2^2 & \cdots & \lambda_2^{N-2}
\\
\vdots & \vdots & \vdots & \ddots &\vdots
\\
 1 & \lambda_{N-1} & \lambda_{N-1}^2 & \cdots & \lambda_{N-1}^{N-2}
\end{pmatrix}
\begin{pmatrix}
F_1
\\
F_2
\\
\vdots
\\
F_{N-1}
\end{pmatrix}
=
\begin{pmatrix}
R_1
\\
R_2
\\
\vdots
\\
R_{N-1}
\end{pmatrix},
 \ee
 where
 \be
F_1=\tilde{\mathcal{I}}, \quad  R_a(\lambda_a, \pi_a ) \equiv -4\prod_{i=1}^{N}{(m_i-\lambda_a)} {\pi^2_a}
 +(-1)^N \sum_{i=1}^N\frac{{ g}^2_i\lambda_a}{m_i-\lambda_a}-g_0(-\lambda_a)^{N-1}.
\ee
%
Hence,
the integrals of motion which are the solutions to this equation may then be expressed via the inverse Vandermonde matrix,
\be
	\label{eq:F_general}
	F_{N-1}=\sum_{i=1}^{N-1}\frac{R_i}{\prod\limits_{\substack{j=1\\j\ne i}}^{N-1}(\lambda_i-\lambda_j)}, \quad
	F_a=(-1)^{a-1}\sum_{i=1}^{N-1}R_i\frac{\sum\limits_{\substack{1\le k_1 < ... < k_{N-a-1}\\k_1,...,k_{N-a-1}\ne i}}^{N-1}\lambda_{k_1}\ ...\ \lambda_{k_{N-a-1}}}{\prod\limits_{\substack{j=1\\j\ne i}}^{N-1}(\lambda_j-\lambda_i)}, \qquad \text{where}\ a=\{1,...,N-2\}
\ee
Note that  $F_1= \widetilde{\mathcal{I}}$ is written in ellipsoidal  coordinates in \eqref{24} and in initial "Cartesian" coordinates in  \eqref{L}. In the rest of the paper we will give the  expressions of remaining integrals of motion  in "Cartesian" coordinates  for the special cases  of $N=3,4,5$.
For this purpose we will need, in particular,
the expression of $\pi_a$ via "Cartesian" coordinates and momenta
\be
\label{eq:pi}
\pi_a=-\frac{1}{2}\sum_{b=1}^{N-1}\frac{p_b}{x_b}\frac{\prod\limits_{\substack{i=1\\{i\ne a}}}^{N}(m_b-\lambda_i)}{\prod\limits_{\substack{i=1\\{i\ne b}}}^{N}(m_b-m_i)}
=-\frac{1}{2}\sum_{b=1}^{N-1}\frac{p_b x_b}{m_b-\lambda_a}
\ee
and the following identity between $N-1$ independent variables $\lambda_a$  which holds for any parameter $\kappa$
\be
\begin{aligned}
	\label{eq:relation}
	\sum_{a=1}^{N-1}\frac{1}{\prod\limits_{\substack{b=1\\{b\ne a}}}^{N-1}(\lambda_a-\lambda_b)}\frac{\lambda_a^\alpha}{\kappa-\lambda_a}=\frac{\kappa^\alpha}{\prod\limits_{a=1}^{N-1}(\kappa-\lambda_a)}-\delta_{\alpha,N-1} \qquad \text{$\alpha=\{0,...,N-1\}$.}
\end{aligned}
\ee
\subsection{$N=3$ case}
For the simplest case of $N=3$, corresponding to seven-dimensional MP black hole, we have two integrals of motion given by \eqref{eq:F_general}:
\be
F_1=\casimir=\frac{\lambda_1R_2- \lambda_2R_1}{\lambda_1-\lambda_2},\qquad F_2=\frac{R_1-R_2}{\lambda_1-\lambda_2}
\ee
Using the expression \eqref{eq:pi} one can explicitly calculate $F_2=F_{N-1}$, which is valid for any $N$ \footnote{Hereinafter, we ignore an additional constant term and  an overall constant factor which might arise in the expressions for the first integrals.}
\be
\label{eq:F_last}
F_{N-1}=\left(\sum_{a=1}^{N-1}p_a x_a\right)^2-\sum_{a=1}^{N-1}p_a^2 m_a-\sum_{i=1}^{N}\frac{m_ip_{\varphi_i}^2}{x_i^2}+g_0\sum_{i=1}^{N}x_i^2\; .
\ee

%

\subsection{$N=4$ case}
	As before, in the case of $N=4$ we will obtain all three integrals of motion from the equation \eqref{eq:F_general}:
	\begin{equation}
	\label{eq:nu0nu1nu2}
	\begin{aligned}
		F_1&=\casimir=\frac{1}{D}\bigg(R_1\lambda_2\lambda_3(\lambda_3-\lambda_2)-R_2\lambda_1\lambda_3(\lambda_3-\lambda_1)+R_3\lambda_1\lambda_2(\lambda_2-\lambda_1)\bigg),\\
		F_2&=\frac{1}{D}\bigg(R_1(\lambda_2^2-\lambda_3^2)+R_2(\lambda_3^2-\lambda_1^2)+R_3(\lambda_1^2-\lambda_2^2)\bigg),\\
		F_3&=\frac{1}{D}\bigg(R_1(\lambda_3-\lambda_2)+R_2(\lambda_1-\lambda_3)+R_3(\lambda_2-\lambda_1)\bigg),
	\end{aligned}
	\end{equation}
	where \[D=\lambda_2\lambda_3(\lambda_3-\lambda_2)-\lambda_1\lambda_3(\lambda_3-\lambda_1)+\lambda_1\lambda_2(\lambda_2-\lambda_1).\]
Constants of motion $F_1$ and $F_3$ in Cartesian coordinates are given by \eqref{L} and \eqref{eq:F_last} respectively.	The second integral of motion can be derived by directly transforming the  second equation in \eqref{eq:nu0nu1nu2}. 
Using \eqref{eq:pi} and \eqref{eq:relation} we derive the expression for  $F_2=F_{N-2}$ which is valid for any $N$
\be
\label{eq:N-2}
F_{N-2}=\sum_{a,b=1}^{N-1}p_a x_a p_b x_b\sum_{\substack{k=1\\k\ne a,b}}^{N}m_k-\sum_{a=1}^{N-1}(p_a x_a)^2m_a+\sum_{a=1}^{N-1}p_a^2(m_a^2-f_1m_a)+\sum_{i=1}^{N}\frac{p_{\varphi_i}^2}{x_i^2}(m_i^2-f_1m_i)+g_0\sum_{\substack{i,j\\i\ne j}}^{N}m_i x_j^2,
\ee
where
\be
\label{eq:f1f2}
f_1(x_i,m_j)\equiv\sum_{i}^{N}(-{x_i}^2+m_i).
\ee
\subsection{$N=5$ case}
In this case we have four integrals of motion, three of which are given by \eqref{L}, \eqref{eq:F_last} and \eqref{eq:N-2}. The missing one is $F_2$ which, as in the previous cases, is given by \eqref{eq:F_general} in ellipsoidal coordinates
\be
	F_2=\sum_{i=1}^{4}R_i
	\frac{\sum\limits_{j=1}^{3}(-1)^{j-1}\lambda_i^{3-j}f_{j-1}}
	{\prod\limits_{\substack{k=1\\k\ne i}}^{4}(\lambda_i-\lambda_k)},
\ee
where $f_0\equiv1$, $f_1$ is given by \eqref{eq:f1f2} and
\be
	f_2(x_i,m_j)\equiv\sum_{\substack{i, j\\i \ne j}}^{N}m_i(-x_j^2+\frac{m_j}{2}).
\ee
Using \eqref{eq:pi} and \eqref{eq:relation} we can represent $F_2=F_{N-3}$ in Cartesian coordinates
and generalize it to higher dimensions
\be
\begin{aligned}
	\label{eq:N-3}
	F_{N-3}=&\sum_{j,k}^{N-1}p_jx_jp_kx_k M_2^{\ne k,j}-\sum_{j=1}^{N-1}(p_jx_j)^2(M_1^{\ne j}m_j-m_j^2)\\
	&-\sum_{j=1}^{N-1}p_j^2(m_jf_2-m_j^2f_1+m_j^3)-\sum_{j=1}^{N}\frac{p_{\varphi_j}^2}{x_j^2}(m_jf_2-m_j^2f_1+m_j^3)+g_0\sum_{i=1}^{N}x_i^2 M_2^{\ne i}
\end{aligned}
\ee
Here we use the following notations
\be
	M_1^{\ne j}\equiv\sum_{\substack{k=1\\k\ne j}}^{N}m_k,\qquad M_2^{\ne j_1,..,j_a}\equiv\sum_{\substack{k_1,k_2=1\\k_1,k_2\ne j_1,..,j_a}}^{N}m_{k_1}m_{k_2}.
\ee
\subsection{Contraction to "isotropic" case}
The expressions for constants of motion  for conformal mechanics associated with $(2N+1)$-dimensional Myers-Perry black holes with non-equal rotational parameters, given by \eqref{eq:F_general}, were initially derived in \cite{non-equal-general} in ellipsoidal coordinates. They were introduced under assumption that all parameters $m_i$ are non-equal and non-zero. In previous subsections we represented them in the initial "Cartesian" coordinates for $N=3,4,5$ cases and found that these expressions are valid in any dimension.
Thus, one can expect that in the "isotropic" limit,  when all parameters $m_i$ are equal to each other ($m_i=N$), we should find a subset of first integrals  of the isotropic Myers-Perry black hole studied in \cite{Anton-MP}, \cite{GNS-1}. However, in this limit, all derived integrals \eqref{L}, \eqref{eq:F_last}, \eqref{eq:N-2} and \eqref{eq:N-3} result in (after rescaling $x_i \to x_i\sqrt{N}$, $p_i\to p_i/\sqrt{N}$)  the  Hamiltonian of a spherical mechanics with equal rotation parameters (with an additional constant term and an overall constant coefficient)
\be
	\label{eq:equal_rot}
	\casimir=\sum_{a,b=1}^{N-1}(\delta_{ab}-x_a x_b)p_ap_b+\sum_{i=1}^{N}\frac{p_{\varphi_i}^2}{x_i^2}, \qquad x_N^2=1-\sum_{a=1}^{N-1}x_a^2.
\ee
This system is not just integrable, but even superintegrable. Thus, naive contruction of the integrals of motion obtained in elliptic coordinates, do not result in the
independent set of Liuvvile integrals for the system with equal rotational parameters. Hence, to obtain the constants of motion in "isotropic" case we need to develop  less straightforward contruction procedure.

Finally, we should note that though we demonstrate this for the $F_{N-1}$, $F_{N-2}, F_{N-3}$, there is no doubt that the same takes place for the rest of the integrals of motion.

\section{Summary and Discussion}
\label{sec:conclusion}
As it was mentioned before the integrability of EMP black hole with nonequal nonvanishing rotation parameters has been proven in \cite{non-equal-general}. In the current paper we have studied the special cases of seven, nine and eleven dimensions. The first integrals of the spherical mechanics in these dimensions have been derived in ellipsoidal and in the initial ``Cartesian'' coordinates. It is important to note, that these results are easily generalized to higher dimensions.

One could assume that in the isotropic limit, if we take all the rotation parameters to be equal to each other, the integrals of motion should transform into the corresponding first integrals of an EMP black hole with equal rotation parameters. But as it turns out, in such a special case all the first integrals become equal to each other and will correspond to the Hamiltonian of the spherical mechanics of an EMP black hole with equal rotation parameters. Hence, one can say that if the rotation parameters are not equal the Hamiltonian of the spherical mechanics of an EMP black hole ``decomposes'' into $N-1$ different first integrals. But when the rotational parameters are equal, additional integrals of motion arise as a result of an enhancement of the rotational symmetry.

A logical development of the problem is to find a general expression for all $N-1$ integrals of motion in Cartesian coordinates in any dimension. This should be done by either generalizing the first integrals \eqref{L}, \eqref{eq:F_last}, \eqref{eq:N-2} and \eqref{eq:N-3} or by directly transforming \eqref{eq:F_general}. Further interesting problem includes the study of EMP black hole with nonequal nonvanishing rotation parameters in even dimensions, which should somehow be related to the solution in odd dimensions. Another important problem is the spacial case of EMP black hole with equal rotation parameters when one of the parameters is equal to zero \cite{evh}. As it turns out, the spherical mechanics of this case does not correspond to the appropriate limiting case of EMP black hole with equal rotation parameters and requires an independent treatment.

	\acknowledgments
I would like to thank Armen Nersessian for suggesting me this problem and for numerous discussions during my work.
This work  was done within ICTP programs NT04 and OEA-AC-100.

\end{document}